# Convolution, Product and Correlation Theorems for Simplified Fractional Fourier Transform: A Mathematical Investigation

### Sanjay Kumar*

*Department of ECE, Thapar Institute of Engineering and Technology, Patiala, Punjab, India*

E-mail: (*-Corresponding Author) *sanjay.kumar@thapar.edu*

_______________________________________________________________________________________

**Abstract**

The notion of fractional Fourier transform (FrFT) has been used and investigated for many years by various research communities, which finds widespread applications in many diverse fields of research study. The potential applications includes ranging from quantum physics, harmonic analysis, optical information processing, pattern recognition to varied allied areas of signal processing. Many significant theorems and properties of the FrFT have been investigated and applied to many signal processing applications, most important among these are convolution, product and correlation theorems. Still many magnificent research works related to the conventional FrFT lacks the elegance and simplicity of the convolution, product and correlation theorems similar to the Euclidean Fourier transform (FT), which for convolution theorem states that the FT of the convolution of two functions is the product of their respective FTs. The purpose of this paper is to devise the equivalent elegancy of convolution, product and correlation theorems, as in the case of Euclidean FT. Building on the seminal work of Pei *et al*. and the potential of the *simplified fractional Fourier transform* (SmFrFT), a detailed mathematical investigation is established to present an elegant definition of convolution, product and correlation theorems in the SmFrFT domain, along with their associated important properties. It has been shown that the established theorems along with their associated properties very nicely generalizes to the classical Euclidean FT.

## 1. Introduction

As it is well-known that the FT is one of the best and most valuable tools used in signal processing and analysis for centuries. It finds its diverse application areas in science and engineering [1, 2]. The fractional Fourier transform (FrFT) is a generalization of the Euclidean Fourier transform (FT), which has found to have several applications in the areas of optics and signal processing [3]. It leads to the generalization of the notion of space (or time) and frequency domains, which are central concepts of signal processing [4-12].

It is defined via an integral as

$$\mathbb{F}^{\varphi}[x(t)] = X^{\varphi}\left(u_{\varphi}\right) = \int_{-\infty}^{\infty} x(t)\, \mathbb{K}_{\varphi}\left(t, u_{\varphi}\right) dt \tag{1}$$

where the transformation kernel $\mathbb{K}_{\varphi}\left(t, u_{\varphi}\right)$ of the FrFT is given by

$$\mathbb{K}_{\varphi}\left(t, u_{\varphi}\right) = \begin{cases} \sqrt{\dfrac{1-j\cot\varphi}{2\pi}} \exp\left[\dfrac{j}{2}\left(u_{\varphi}^{2} + t^{2}\right)\cot\varphi - ju_{\varphi}t\csc\varphi\right] & \text{if } \varphi \neq n\pi, (n = 0,1,2,\cdots) \\ \delta\left(t - u_{\varphi}\right) & \text{if } \varphi = 2n\pi, \\ \delta\left(t + u_{\varphi}\right) & \text{if } \varphi = (2n+1)\pi \end{cases} \tag{2}$$

where $\varphi$ indicates the rotation angle of the transformed signal in the FrFT domain.

When $\varphi = \pi/2$, the FrFT reduces to the Euclidean FT, and when $\varphi = 0$, it is the same as the identity operation. It also satisfies the additivity property $\mathbb{F}^{\varphi}\{\mathbb{F}^{\gamma}[x(t)]\} = \mathbb{F}^{\gamma}\{\mathbb{F}^{\varphi}[x(t)]\} = \mathbb{F}^{\varphi+\gamma}[x(t)]$. The detailed properties were described in [3,4,6]. Thus, it is a generalization of the Euclidean FT and is regarded as a counter-clockwise rotation of the signal coordinates around the origin in the time-frequency (TF) plane with the rotation angle $\varphi$ [3, 9, 10, 12].

As it is well-known that the FrFT is able to process non-stationary or chirp signals better than the Euclidean FT, due to the reason that a chirp signal forms a line or an impulse in the TF plane and thus there exists a fractional transformation order in which such signals are compact [13]. Chirp signals are not compact in the time or spatial domain. Thus, one can extract the signal easily in an appropriate (optimum) fractional Fourier domain, when it is not possible to separate the signal and noise in the spatial or Fourier frequency domain. It is this introduction of extra degree of freedom which gives the FrFT a potential improvement over the Euclidean FT [3].

In [14], the authors have introduced various simplified forms of the FrFT, known as the *simplified fractional Fourier transform* (SmFrFT). The reason for establishing the SmFrFT is that they are simplest for digital computation, optical implementation, graded-index (GRIN) medium implementation, and radar system implementation, with the same capabilities as the conventional FrFT for designing fractional filters or for fractional correlation. The SmFrFT possess a great potential for replacing the conventional FrFT in many applications. Pei *et al.* [14] establishes five types of SmFrFTs that have the same capabilities as the conventional FrFT for the fractional filter design or for fractional correlation and simultaneously are simplest for digital computation, optical implementation, and radar system implementation. Thus, SmFrFTs have a great potential to substitute for the conventional FrFTs in many real-time applications [12, 15-18]. Another dominant advantage that SmFrFT possess over the conventional FrFT is its less computational complexity, as is evident from [14].

In this paper, the main focus is on Type 1 SmFrFT, because it is simpler for digital implementation, which was discussed in [14]. Many properties of the FrFT are currently well-known [19-21], including its convolution, product and correlation theorems. However, the convolution, product and correlation theorems for the conventional FrFT [19-21] do not generalize the classical result of the Euclidean FT [1, 2]. As it well-known that the convolution theorem of the Euclidean FT for the functions $f(t)$ and $g(t)$ with associated Euclidean FTs, $F(\omega)$ and $G(\omega)$, respectively is given by [1, 2]

$$f(t) \circledast g(t) \overset{\mathcal{F}}{\leftrightarrow} F(\omega)G(\omega) \tag{3}$$

where $\mathcal{F}$ and $\circledast$ denotes the Euclidean FT operation and the convolution operation, respectively. This convolution theorem can also be written as [1, 2]

$$\int_{-\infty}^{\infty} f(\tau)g(t-\tau)d\tau \overset{\mathcal{F}}{\leftrightarrow} F(\omega)G(\omega) \tag{4}$$

Thus, the convolution theorem states that the convolution of two time-domain functions results in simple multiplication of their Euclidean FTs in the Euclidean FT domain—a really powerful result. Similar is the case with correlation theorem in the Euclidean FT domain for two complex-valued functions, which is given by [1, 2]

$$\bar{f}(t) \circledcirc g(t) \overset{\mathcal{F}}{\leftrightarrow} \bar{F}(-\omega)G(\omega) \tag{5}$$

where $\mathcal{F}$, $\circledcirc$ and $\overline{(\cdot)}$ denotes the Euclidean FT operation, the correlation operation, and the complex conjugate, respectively. This correlation theorem can also be written as [1, 2]

$$\int_{-\infty}^{\infty} \bar{f}(\tau)g(\tau+t)d\tau \overset{\mathcal{F}}{\leftrightarrow} \bar{F}(-\omega)G(\omega) \tag{6}$$

Thus, the correlation theorem states that multiplying the Euclidean FT of one function with the complex conjugate of the other function gives the Euclidean FT of their correlation.

However, the convolution and correlation theorem for the conventional FrFT lacks this simplicity and elegancy in the analytical result [19-21]. Various researchers in the fractional signal processing society [22-27] have developed different modified versions of these theorems in the FrFT and linear canonical transform (LCT) domains, by utilizing the conventional definition of their transforms. But, still there exists a room for its improvement to reflect upon the elegancy and the simplicity of the theorems. In this paper, convolution, product and correlation theorems are proposed based the simplified FrFT, which preserves the elegance and simplicity comparable to that of the Euclidean FT, which finds widespread applications in various allied research areas of signal processing [15-17]. The conventional convolution, product and correlation theorems and its associated properties are shown to be special cases of the derived results.

The rest of the paper is organized as follows. In Section 2, the preliminaries of the simplified fractional Fourier transform (SmFrFT) is presented, followed by the convolution, product and the correlation theorems associated with SmFrFT in Sections 3, 4 and 5, respectively, along with the analytical proofs of their associated properties. Finally, conclusions and the future scope of work for the fractional signal processing society is summarized in Section 5.

## 2 Preliminaries

### 2.1 The Simplified Fractional Fourier Transform: Definition and Integral Representation

The simplified fractional Fourier transform (SmFrFT) of the signal $x(t)$ is represented by [14]:

$$T_S^\varphi[x(t)] = X_S^\varphi(u_\varphi) = \int_{-\infty}^{\infty} x(t)\, K_\varphi^S(t, u_\varphi) dt \qquad (7)$$

where $K_\varphi^S(t, u_\varphi) = \frac{1}{\sqrt{j2\pi}} \exp\left[-j\, t\, u_\varphi + \frac{j}{2} t^2 \cot \varphi\right]$ \qquad (8)

Here, $t$ and $u_\varphi$ can interchangeably represent time and fractional frequency domains. The transform output lies between time and frequency domains, except for the special cases of $\varphi = 0$ and $\varphi = \pi/2$ which belongs to FT domain. Based upon (7), the SmFrFT can be realized in a three step process (see Fig. 1), as opposed to the conventional FrFT which is realized by a four step process, as follows:

(i) pre-multiplication of the input signal by a linear chirp with the frequency modulation (FM) rate determined by the fractional rotation angle $\varphi$ or the fractional transformation order $a$, related by $\varphi = a\pi/2$ with $a \in \mathbb{R}$;

(ii) computation of the Euclidean FT ($\mathcal{F}$);

(iii) post-multiplication by a complex amplitude factor.

The inverse SmFrFT is given by [14]:

$$x(t) = \sqrt{\frac{j}{2\pi}}\, e^{-\frac{j}{2} t^2 \cot \varphi} \int_{-\infty}^{\infty} e^{j u_\varphi t}\, X_S^\varphi(u_\varphi)\, du_\varphi \qquad (9)$$

where $X_S^\varphi(u_\varphi)$ represents the SmFrFT of the input signal $x(t)$.

## 3 Convolution Theorem associated with SmFrFT

**Theorem 3.1** *For any two functions $f, g \in L^1(\mathbb{R})$, let $F_S^\varphi, G_S^\varphi$ denote the SmFrFT of $f, g$, respectively. The convolution operator of the SmFrFT is defned as*

$$(f \circledast g)(t) = \int_{-\infty}^{\infty} f(\tau) \, g(t-\tau) \, W_{cv}(\tau, t) \, d\tau \qquad (10)$$

where, $W_{cv}(\tau, t) = e^{j\tau(\tau-t)\cot\varphi}$. Then, the SmFrFT of the convolution of two complex functions is given by

$$\mathbb{T}_S^{\varphi}\{f \circledast g\}(u_{\varphi}) = \sqrt{j2\pi} \, F_S^{\varphi}(u_{\varphi}) \, G_S^{\varphi}(u_{\varphi}) \qquad (11)$$

**Proof:** From the definition of SmFrFT (7) and the SmFrFT convolution (10), one obtains

$$\mathbb{T}_S^{\varphi}\{f \circledast g\}(u_{\varphi}) = \frac{1}{\sqrt{j2\pi}} \int_{-\infty}^{\infty} \{f(t) \circledast g(t)\} e^{-j\,t\,u_{\varphi}+\frac{j}{2}t^2\cot\varphi} dt \qquad (12)$$

$$\mathbb{T}_S^{\varphi}\{f \circledast g\}(u_{\varphi}) = \frac{1}{\sqrt{j2\pi}} \int_{-\infty}^{\infty} \{\int_{-\infty}^{\infty} f(\tau) \, g(t-\tau) \, W_{cv}(\tau, t) \, d\tau\} e^{-j\,t\,u_{\varphi}+\frac{j}{2}t^2\cot\varphi} dt \qquad (13)$$

For solving (13), letting $(t - \tau) = \zeta$

$$\mathbb{T}_S^{\varphi}\{f \circledast g\}(u_{\varphi}) = \frac{1}{\sqrt{j2\pi}} \int_{-\infty}^{\infty} \int_{-\infty}^{\infty} f(\tau) \, g(\zeta) \, e^{-j\tau\zeta\cot\varphi} e^{-j\,(\zeta+\tau)\,u_{\varphi}+\frac{j}{2}(\zeta+\tau)^2\cot\varphi} d\tau \, d\zeta \qquad (14)$$

Rearranging and multiplying numerator and denominator of (14) by $1/\sqrt{2\pi}$, one obtains

$$\mathbb{T}_S^{\varphi}\{f \circledast g\}(u_{\varphi}) = \frac{1}{\sqrt{j2\pi}} \int_{-\infty}^{\infty} f(\tau) \, e^{-ju_{\varphi}\tau+\frac{j}{2}\tau^2\cot\varphi} \, d\tau \times \frac{1}{\sqrt{j2\pi}} \int_{-\infty}^{\infty} g(\zeta) \, e^{-ju_{\varphi}\zeta+\frac{j}{2}\zeta^2\cot\varphi} \, d\zeta \times \sqrt{j2\pi} \qquad (15)$$

By the definition of SmFrFT, the above expression (15) reduces to

$$\mathbb{T}_S^{\varphi}\{f \circledast g\}(u_{\varphi}) = \sqrt{j2\pi} \, F_S^{\varphi}(u_{\varphi}) \, G_S^{\varphi}(u_{\varphi}), \qquad (16)$$

which proves the theorem in SmFrFT domain.

**Special case:**

For the Euclidean FT, the rotation angle $\varphi = \pi/2$, then the expression (16) reduces to

$$\mathbb{T}_S^{\pi/2}\{f \circledast g\}(u_{\pi/2}) = \sqrt{j2\pi} \, F_S^{\pi/2}(u_{\pi/2}) \, G_S^{\pi/2}(u_{\pi/2}) = \sqrt{j2\pi} \, F_S(\omega) \, G_S(\omega) \qquad (17)$$

This means that the proposed convolution theorem behaves similar to the Euclidean FT, i.e., the convolution in the time-domain is equivalent to the multiplication in the simplified fractional frequency domain, except for the amplitude factor $\sqrt{j2\pi}$ and where $u_{\pi/2} = \omega$.

Some properties associated with the convolution theorem in SmFrFT domain are illustrated below.

***Property 1*** (**Shift Convolution**). *Let* $f, g \in L^1(\mathbb{R})$. *The SmFrFT of* $\mathbb{S}_d f \circledast g$ *and* $f \circledast \mathbb{S}_d g$ *is given by*

$$\mathbb{T}_S^\varphi \{\mathbb{S}_d f \circledast g\}(u_\varphi) = \sqrt{j2\pi}\, e^{-ju_\varphi d + \frac{j}{2} d^2 \cot \varphi} F_S^\varphi (u_\varphi - d \cot \varphi)\, G_S^\varphi (u_\varphi) \tag{18}$$

$$\mathbb{T}_S^\varphi \{f \circledast \mathbb{S}_d g\}(u_\varphi) = \sqrt{j2\pi}\, e^{-ju_\varphi d + \frac{j}{2} d^2 \cot \varphi} F_S^\varphi (u_\varphi)\, G_S^\varphi (u_\varphi - d \cot \varphi) \tag{19}$$

where, the symbol $\mathbb{S}_d$ represents the shift operator of a function by delay $d$ i.e., $\mathbb{S}_d x(t) = x(t - d)$, $d \in \mathbb{R}$.

***Proof:*** The shift convolution operator $\mathbb{S}_d f \circledast g$ is given by

$$(\mathbb{S}_d f \circledast g)(t) = \int_{-\infty}^\infty f(\tau - d)\, g(t - \tau)\, W_{cv}(\tau, t)\, d\tau \tag{20}$$

where, $W_{cv}(\tau, t) = e^{j\tau(\tau - t)\cot \varphi}$. It implies

$$(\mathbb{S}_d f \circledast g)(t) = \int_{-\infty}^\infty f(\tau - d)\, g(t - \tau)\, e^{j\tau(\tau - t)\cot \varphi}\, d\tau \tag{21}$$

Now, from the definition of SmFrFT (7), one obtains

$$\mathbb{T}_S^\varphi \{\mathbb{S}_d f \circledast g\}(u_\varphi) = \frac{1}{\sqrt{j2\pi}} \int_{-\infty}^\infty \{\mathbb{S}_d f(t) \circledast g(t)\}\, e^{-j\, t\, u_\varphi + \frac{j}{2} t^2 \cot \varphi}\, dt \tag{22}$$

Simplifying (22) further, one obtains

$$\mathbb{T}_S^\varphi \{\mathbb{S}_d f \circledast g\}(u_\varphi) = \frac{1}{\sqrt{j2\pi}} \int_{-\infty}^\infty \left\{ \int_{-\infty}^\infty f(\tau - d)\, g(t - \tau)\, e^{j\tau(\tau - t)\cot \varphi}\, d\tau \right\} e^{-j\, t\, u_\varphi + \frac{j}{2} t^2 \cot \varphi}\, dt \tag{23}$$

To solve (23), let's assume $(t - \tau) = p$

$$\mathbb{T}_S^\varphi \{\mathbb{S}_d f \circledast g\}(u_\varphi) = \frac{1}{\sqrt{j2\pi}} \int_{-\infty}^\infty f(\tau - d)\, e^{-ju_\varphi \tau + \frac{j}{2}\tau^2 \cot \varphi}\, d\tau \times \int_{-\infty}^\infty g(p)\, e^{-ju_\varphi p + \frac{j}{2} p^2 \cot \varphi}\, dp$$

$$\mathbb{T}_S^\varphi \{\mathbb{S}_d f \circledast g\}(u_\varphi) = \int_{-\infty}^\infty f(\tau - d)\, e^{-ju_\varphi \tau + \frac{j}{2}\tau^2 \cot \varphi}\, d\tau \times G_S^\varphi (u_\varphi) \tag{24}$$

Further, by letting $(\tau - d) = z$, and multiplying numerator and denominator of () by $\sqrt{j2\pi}$, one solves (24) as

$$\mathbb{T}_S^\varphi\{\mathbb{S}_d f \circledast g\}(u_\varphi) = \frac{1}{\sqrt{j2\pi}}\, e^{-ju_\varphi d + \frac{j}{2}d^2\cot\varphi} \int_{-\infty}^{\infty} f(z)\, e^{-j(u_\varphi - d\cot\varphi)z + \frac{j}{2}z^2\cot\varphi} dz \;\times\; G_S^\varphi(u_\varphi) \times \sqrt{j2\pi}$$

$$\mathbb{T}_S^\varphi\{\mathbb{S}_d f \circledast g\}(u_\varphi) = \sqrt{j2\pi}\, e^{-ju_\varphi d + \frac{j}{2}d^2\cot\varphi} F_S^\varphi(u_\varphi - d\cot\varphi)\, G_S^\varphi(u_\varphi),$$

which proves the shift convolution property.

Similarly, for solving $\mathbb{T}_S^\varphi\{f \circledast \mathbb{S}_d g\}(u_\varphi)$ and utilizing the shift convolution operator of function $f \circledast \mathbb{S}_d g$ as $\int_{-\infty}^{\infty} f(\tau)\, g(t - \tau - d)\, W_{cv}(\tau, t)\, d\tau$, where, $W_{cv}(\tau, t) = e^{j\tau(\tau - t)\cot\varphi}$ and based on the previous steps, one obtains

$$\mathbb{T}_S^\varphi\{f \circledast \mathbb{S}_d g\}(u_\varphi) = \sqrt{j2\pi}\, e^{-ju_\varphi d + \frac{j}{2}d^2\cot\varphi} F_S^\varphi(u_\varphi)\, G_S^\varphi(u_\varphi - d\cot\varphi), \tag{25}$$

which proves the shift convolution property in SmFrFT domain.

Thus, (24) and (25) indicates that if we apply a linear time delay to one signal in the time domain and convolve it with the another time domain signal, then the SmFrFT of the convolved signal is identical to the multiplications of the SmFrFTs of the respective signals, except that one of the signal has been shifted in the SmFrFT domain by an amount dependent on the change in time shift in the time domain, and there is a multiplication with the complex harmonic dependent on the time shift.

***Special case*:**

For the Euclidean FT, the rotation angle $\varphi = \pi/2$, then the expression (24) and (25) reduces to

$$\mathbb{T}_S^{\pi/2}\{\mathbb{S}_d f \circledast g\}(u_{\pi/2}) = \sqrt{j2\pi}\, e^{-ju_{\pi/2}d} F_S^{\pi/2}(u_{\pi/2})\, G_S^{\pi/2}(u_{\pi/2})$$

i.e, $\mathcal{F}\{\mathbb{S}_d f \circledast g\}(\omega) = \sqrt{j2\pi}\; e^{-j\omega d}\, F_S(\omega)\, G_S(\omega)$ (26)

$$\mathbb{T}_S^{\pi/2}\{f \circledast \mathbb{S}_d g\}(u_{\pi/2}) = \sqrt{j2\pi}\, e^{-ju_{\pi/2}d} F_S^{\pi/2}(u_{\pi/2})\, G_S^{\pi/2}(u_{\pi/2})$$

i.e, $\mathcal{F}\{f \circledast \mathbb{S}_d g\}(\omega) = \sqrt{j2\pi}\; e^{-j\omega d}\, F_S(\omega)\, G_S(\omega)$ (27)

This means that the proposed shift convolution property behaves similar to the Euclidean FT, as is evident from (26) and (27), respectively.

***Property 2*** (**Modulation Convolution**). *Let* $f, g \in L^1(\mathbb{R})$. *The SmFrFT of* $\mathbb{M}_q f \circledast g$ *and* $f \circledast \mathbb{M}_q g$ *is given by*

$$\mathbb{T}_S^{\varphi}\{\mathbb{M}_q f \circledast g\}(u_{\varphi}) = \sqrt{j2\pi} \, F_S^{\varphi}(u_{\varphi} - q) \, G_S^{\varphi}(u_{\varphi}) \tag{28}$$

$$\mathbb{T}_S^{\varphi}\{f \circledast \mathbb{M}_q g\}(u_{\varphi}) = \sqrt{j2\pi} \, F_S^{\varphi}(u_{\varphi}) \, G_S^{\varphi}(u_{\varphi} - q) \tag{29}$$

where, the symbol $\mathbb{M}_q$ represents the modulation operator, i.e., the modulation by $q$ of a function $x(t)$, $\mathbb{M}_q x(t) = e^{jqt} x(t), q \in \mathbb{R}$.

***Proof:*** The modulation convolution operator $\mathbb{M}_q f \circledast g$ is given by

$$\left(\mathbb{M}_q f \circledast g\right)(t) = \int_{-\infty}^{\infty} e^{jq\tau} f(\tau) \, g(t - \tau) \, W_{cv}(\tau, t) \, d\tau \tag{30}$$

where, $W_{cv}(\tau, t) = e^{j\tau(\tau - t)\cot\varphi}$. It implies

$$\left(\mathbb{M}_q f \circledast g\right)(t) = \int_{-\infty}^{\infty} e^{jq\tau} f(\tau) \, g(t - \tau) \, e^{j\tau(\tau - t)\cot\varphi} \, d\tau \tag{31}$$

Now, from the definition of SmFrFT (7), one obtains

$$\mathbb{T}_S^{\varphi}\{\mathbb{M}_q f \circledast g\}(u_{\varphi}) = \frac{1}{\sqrt{j2\pi}} \int_{-\infty}^{\infty} \{\mathbb{M}_q f(t) \circledast g(t)\} \, e^{-j \, t \, u_{\varphi} + \frac{j}{2}t^2 \cot\varphi} dt \tag{32}$$

Simplifying (32) further, one obtains

$$\mathbb{T}_S^{\varphi}\{\mathbb{M}_q f \circledast g\}(u_{\varphi}) = \frac{1}{\sqrt{j2\pi}} \int_{-\infty}^{\infty} \int_{-\infty}^{\infty} f(\tau) \, g(t - \tau) \, e^{jq\tau + j\tau(\tau - t)\cot\varphi - ju_{\varphi}t + \frac{j}{2}t^2 \cot\varphi} d\tau \, dt \tag{33}$$

By letting $(t - \tau) = v$, (33) reduces to

$$\mathbb{T}_S^{\varphi}\{\mathbb{M}_q f \circledast g\}(u_{\varphi}) = \frac{1}{\sqrt{j2\pi}} \int_{-\infty}^{\infty} f(\tau) \, e^{-j(u_{\varphi} - q)\tau + \frac{j}{2}\tau^2 \cot\varphi} d\tau \times \frac{1}{\sqrt{j2\pi}} \int_{-\infty}^{\infty} h(v) \, e^{-ju_{\varphi}v + \frac{j}{2}v^2 \cot\varphi} dv \times \sqrt{j2\pi}$$

Simplifying further, one obtains

$$\mathbb{T}_S^\varphi \{ \mathbb{M}_q f \circledcirc g \}(u_\varphi) = \sqrt{j2\pi} \, F_S^\varphi(u_\varphi - q) \, G_S^\varphi(u_\varphi), \tag{34}$$

which proves the modulation convolution property in SmFrFT domain.

Similarly, for solving $\mathbb{T}_S^\varphi \{ f \circledcirc \mathbb{M}_q g \}(u_\varphi)$ and utilizing the modulation convolution operator of function $f \circledcirc \mathbb{M}_q g$ as $\int_{-\infty}^\infty f(\tau) \, e^{jq(t-\tau)} \, g(t-\tau) \, W_{cv}(\tau, t) \, d\tau$, where, $W_{cv}(\tau, t) = e^{j\tau(\tau-t)\cot\varphi}$ and based on the previous steps, one obtains

$$\mathbb{T}_S^\varphi \{ f \circledcirc \mathbb{M}_q g \}(u_\varphi) = \sqrt{j2\pi} \, F_S^\varphi(u_\varphi) \, G_S^\varphi(u_\varphi - q), \tag{35}$$

which proves the modulation convolution property in SmFrFT domain.

Thus, (34) and (35) indicates that if we apply a linear change in phase to one signal in the time domain and convolve it with the another time domain signal, then the SmFrFT of the convolved signal is identical to the multiplications of the SmFrFTs of the respective signals, except that one of the signal has been shifted in the SmFrFT domain by an amount dependent on the change in phase in the time domain.

***Special case:***

In case of FT, (34) and (35) reduces to (for $\varphi = \pi/2$)

$$\mathbb{T}_S^{\pi/2} \{ \mathbb{M}_q f \circledcirc g \}(u_{\pi/2}) = \sqrt{j2\pi} \, F_S^{\pi/2}(u_{\pi/2} - q) \, G_S^{\pi/2}(u_{\pi/2}) = \sqrt{j2\pi} \, F_S(\omega - q) \, G_S(\omega),$$

i.e., $\mathcal{F}\{ \mathbb{M}_q f \circledcirc g \}(\omega) = \sqrt{j2\pi} \, F_S(\omega - q) \, G_S(\omega)$ \hfill (36)

$$\mathbb{T}_S^{\pi/2} \{ f \circledcirc \mathbb{M}_q g \}(u_{\pi/2}) = \sqrt{j2\pi} \, F_S^{\pi/2}(u_{\pi/2}) \, G_S^{\pi/2}(u_{\pi/2} - q)$$

$$\mathcal{F}\{ f \circledcirc \mathbb{M}_q g \}(\omega) = \sqrt{j2\pi} \, F_S(\omega) \, G_S(\omega - q) \tag{37}$$

This means that the proposed modulation convolution property behaves similar to the Euclidean FT, as is evident from (36) and (37), respectively.

***Property 3*** (**Time-Frequency shift Convolution**). *Let $f, g \in L^1(\mathbb{R})$. The SmFrFT of $\mathbb{M}_q \mathbb{S}_d f \circledcirc g$ and $f \circledcirc \mathbb{M}_q \mathbb{S}_d g$ is given by*

$$\mathbb{T}_S^\varphi\{\mathbb{M}_q \mathbb{S}_d f \circledcirc g\}(u_\varphi) = \sqrt{j2\pi}\; e^{-j(u_\varphi - q)d + \frac{j}{2}d^2 \cot\varphi}\; F_S^\varphi(u_\varphi - q - d\cot\varphi)\; G_S^\varphi(u_\varphi) \quad (38)$$

$$\mathbb{T}_S^\varphi\{f \circledcirc \mathbb{M}_q \mathbb{S}_d g\}(u_\varphi) = \sqrt{j2\pi}\; e^{-j(u_\varphi - q)d + \frac{j}{2}d^2 \cot\varphi}\; F_S^\varphi(u_\varphi)\; G_S^\varphi(u_\varphi - q - d\cot\varphi) \quad (39)$$

where, the symbol $\mathbb{S}_d$ and $\mathbb{M}_q$ represents the shift operator of a function by delay $d$ and the modulation operator of a function by $q$, i.e., for the function $x(t)$, $\mathbb{S}_d x(t) = x(t - d)$, $d \in \mathbb{R}$ and $\mathbb{M}_q x(t) = e^{jqt} x(t)$, $q \in \mathbb{R}$.

***Proof:*** The time-frequency shift convolution operator is given by

$$\big(\mathbb{M}_q \mathbb{S}_d f \circledcirc g\big)(t) = \int_{-\infty}^{\infty} e^{jq\tau} f(\tau - d)\, g(t - \tau)\, W_{cv}(\tau, t)\, d\tau \quad (40)$$

where, $W_{cv}(\tau, t) = e^{j\tau(\tau - t)\cot\varphi}$. It implies

$$\big(\mathbb{M}_q \mathbb{S}_d f \circledcirc g\big)(t) = \int_{-\infty}^{\infty} e^{jq\tau} f(\tau - d)\, g(t - \tau)\, e^{j\tau(\tau - t)\cot\varphi}\, d\tau \quad (41)$$

The SmFrFT of (41) is obtained as

$$\mathbb{T}_S^\varphi\{\mathbb{M}_q \mathbb{S}_d f \circledcirc g\}(u_\varphi) = \frac{1}{\sqrt{j2\pi}} \int_{-\infty}^{\infty} \{\mathbb{M}_q \mathbb{S}_d f(t) \circledcirc g(t)\} e^{-j\, t\, u_\varphi + \frac{j}{2}t^2 \cot\varphi}\, dt \quad (42)$$

Simplifying (42) further, one obtains

$$\mathbb{T}_S^\varphi\{\mathbb{M}_q \mathbb{S}_d f \circledcirc g\}(u_\varphi) = \frac{1}{\sqrt{j2\pi}} \int_{-\infty}^{\infty} \int_{-\infty}^{\infty} f(\tau - d)\, g(t - \tau)\, e^{jq\tau + j\tau(\tau - t)\cot\varphi - j\, t\, u_\varphi + \frac{j}{2}t^2 \cot\varphi}\, d\tau\, dt \quad (43)$$

By letting $(t - \tau) = \varsigma$, (43) is simplified as

$$\mathbb{T}_S^\varphi\{\mathbb{M}_q \mathbb{S}_d f \circledcirc g\}(u_\varphi) = \frac{1}{\sqrt{j2\pi}} \int_{-\infty}^{\infty} f(\tau - d)\; e^{jq\tau - ju_\varphi\tau + \frac{j}{2}\tau^2 \cot\varphi}\, d\tau \times \frac{1}{\sqrt{j2\pi}} \int_{-\infty}^{\infty} g(\varsigma)\, e^{-ju_\varphi\varsigma + \frac{j}{2}\varsigma^2 \cot\varphi}\, d\varsigma \times$$
$$\sqrt{j2\pi} \quad (44)$$

Let $(\tau - d) = \xi$, (44) reduces to

$$\mathbb{T}_S^\varphi\{\mathbb{M}_q \mathbb{S}_d f \circledcirc g\}(u_\varphi) = \frac{1}{\sqrt{j2\pi}} \int_{-\infty}^{\infty} f(\xi)\; e^{-j(u_\varphi - q - d\cot\varphi)\xi + \frac{j}{2}\xi^2 \cot\varphi}\, d\xi \; \times \sqrt{j2\pi} \times$$
$$e^{-ju_\varphi d + jqd + \frac{j}{2}d^2 \cot\varphi} \times G_S^\varphi(u_\varphi) \quad (45)$$

Thus, $\mathbb{T}_S^{\varphi}\{\mathbb{M}_q\mathbb{S}_d\,f\,\circledast\,g\,\}(u_\varphi) = \sqrt{j2\pi}\,e^{-j(u_\varphi-q)d+\frac{j}{2}d^2\cot\varphi}\,F_S^{\varphi}(u_\varphi-q-d\cot\varphi)\,G_S^{\varphi}(u_\varphi),$ (46)

which proves the time-frequency shift convolution property in SmFrFT domain.

Similarly, for solving $\mathbb{T}_S^{\varphi}\{f\,\circledast\,\mathbb{M}_q\mathbb{S}_d g\}(u_\varphi)$ and utilizing the shift and modulation convolution operator of function $f\,\circledast\,\mathbb{M}_q\mathbb{S}_d g$ as $\int_{-\infty}^{\infty}f(\tau)\,e^{jq(t-\tau)}\,g(t-\tau-d)\,W_{cv}(\tau,t)\,d\tau$, where, $W_{cv}(\tau,t) = e^{j\tau(\tau-t)\cot\varphi}$ and based on the previous steps, one obtains

$\mathbb{T}_S^{\varphi}\{f\,\circledast\,\mathbb{M}_q\mathbb{S}_d g\}(u_\varphi) = \sqrt{j2\pi}\,e^{-j(u_\varphi-q)d+\frac{j}{2}d^2\cot\varphi}\,F_S^{\varphi}(u_\varphi)\,G_S^{\varphi}(u_\varphi-q-d\cot\varphi),$ (47)

which proves the time-frequency shift convolution property in SmFrFT domain.

***Special case***:

In case of FT, (46) and (47) reduces to (for $\varphi = \pi/2$)

$\mathbb{T}_S^{\pi/2}\{\mathbb{M}_q\mathbb{S}_d\,f\,\circledast\,g\,\}(u_{\pi/2}) = \sqrt{j2\pi}\,e^{-j(u_{\pi/2}-q)d}\,F_S^{\pi/2}(u_{\pi/2}-q)\,G_S^{\pi/2}(u_{\pi/2}),$

i.e., $\mathcal{F}\{\mathbb{M}_q\mathbb{S}_d\,f\,\circledast\,g\,\}(\omega) = \sqrt{j2\pi}\,e^{-j(\omega-q)d}\,F_S(\omega-q)\,G_S(\omega)$ (48)

$\mathbb{T}_S^{\pi/2}\{f\,\circledast\,\mathbb{M}_q\mathbb{S}_d\,g\,\}(u_{\pi/2}) = \sqrt{j2\pi}\,e^{-j(u_{\pi/2}-q)d}\,F_S^{\pi/2}(u_{\pi/2})\,G_S^{\pi/2}(u_{\pi/2}-q)$

i.e, $\mathcal{F}\{f\,\circledast\,\mathbb{M}_q\mathbb{S}_d\,g\,\}(\omega) = \sqrt{j2\pi}\,e^{-j(\omega-q)d}\,F_S(\omega)\,G_S(\omega-q)$ (49)

This means that the proposed time-frequency shift convolution property behaves similar to the Euclidean FT, as is evident from (48) and (49), respectively.

## 4. Product Theorem associated with SmFrFT

**Theorem 4.1** *For any two functions $f$, $g \in L^1(\mathbb{R})$, let $F_S^{\varphi}$, $G_S^{\varphi}$ denote the SmFrFT of $f$, $g$, respectively. We define the product operation associated with SmFrFT as*

$$z(t) = f(t)\,g(t)\,W_p(t) = f(t)\,g(t)\,e^{\frac{j}{2}t^2\cot\varphi} \qquad (50)$$

*where, $W_p(t) = e^{\frac{j}{2}t^2\cot\varphi}$. Then, the SmFrFT of the product of two functions is given by*

$$\mathbb{T}_S^{\varphi}\{z(t)\} = \mathbb{T}_S^{\varphi}\{f(t)\ g(t)\ W_p(t)\} = \sqrt{\frac{j}{2\pi}}\ F_S^{\varphi}(u_{\varphi}) * G_S^{\varphi}(u_{\varphi}) \tag{51}$$

***Proof:*** The function $z$ is in $L^1(\mathbb{R})$ and thus its SmFrFT is given by (7). To compute $\mathbb{T}_S^{\varphi}\{z(t)\}$, express the function $z(t)$ in terms of its SmFrFT, as follows:

$$\mathbb{T}_S^{\varphi}\{z(t)\} = Z_S^{\varphi}(u_{\varphi}) = \frac{1}{\sqrt{j2\pi}}\int_{-\infty}^{\infty}\{f(t)\ g(t)\ W_p(t)\}e^{-j\ t\ u_{\varphi}+\frac{j}{2}t^2\cot\varphi}dt =$$

$$\frac{1}{\sqrt{j2\pi}}\int_{-\infty}^{\infty}\{f(t)\}\ g(t)\ W_p(t)\ e^{-j\ t\ u_{\varphi}+\frac{j}{2}t^2\cot\varphi}dt \tag{52}$$

$$\mathbb{T}_S^{\varphi}\{z(t)\} = Z_S^{\varphi}(u_{\varphi}) = \frac{1}{\sqrt{j2\pi}}\int_{-\infty}^{\infty}e^{-j\ t\ u_{\varphi}+\frac{j}{2}t^2\cot\varphi}\left\{\sqrt{\frac{j}{2\pi}}\ e^{-\frac{j}{2}t^2\cot\varphi}\int_{-\infty}^{\infty}e^{jv_{\varphi}t}\ F_S^{\varphi}(v_{\varphi})\ dv_{\varphi}\right\}g(t)\ W_p(t)\ dt$$

Simplifying further and using $W_p(t) = e^{\frac{j}{2}t^2\cot\varphi}$, one obtains

$$\mathbb{T}_S^{\varphi}\{z(t)\} = Z_S^{\varphi}(u_{\varphi}) = \frac{1}{\sqrt{j2\pi}}\int_{-\infty}^{\infty}e^{-j(u_{\varphi}-v_{\varphi})t+\frac{j}{2}t^2\cot\varphi}\ g(t)\ dt\ \times\ \sqrt{\frac{j}{2\pi}}\int_{-\infty}^{\infty}F_S^{\varphi}(v_{\varphi})\ dv_{\varphi} \tag{53}$$

Solving (53), one gets the following analytical expression

$$\mathbb{T}_S^{\varphi}\{z(t)\} = Z_S^{\varphi}(u_{\varphi}) = \sqrt{\frac{j}{2\pi}}\int_{-\infty}^{\infty}F_S^{\varphi}(v_{\varphi})\ G_S^{\varphi}(u_{\varphi}-v_{\varphi})\ dv_{\varphi} = \sqrt{\frac{j}{2\pi}}\ F_S^{\varphi}(u_{\varphi}) * G_S^{\varphi}(u_{\varphi}), \tag{54}$$

which proves the product theorem in SmFrFT domain.

Thus, the product theorem in SmFrFT domain states that the SmFrFT of the product of two functions is obtained by conventional convolution between the SmFrFTs of the two functions.

## 5. Correlation Theorem associated with SmFrFT

**Theorem 5.1** *For any two complex-valued functions $f$, $g \in L^1(\mathbb{R})$, let $F_S^{\varphi}$, $G_S^{\varphi}$ denote the SmFrFT of $f$, $g$, respectively. We define the correlation operator of the SmFrFT as*

$$(f \circledcirc g)(t) = \int_{-\infty}^{\infty}\bar{f}(\tau)\ g(t+\tau)\ W_{cr}(\tau,t)\ d\tau \tag{55}$$

where, $W_{cr}(\tau, t) = e^{j\tau(\tau+t)\cot\varphi}$. *Then, the SmFrFT of the correlation of the two complex-valued functions is given by*

$$\mathbb{T}_S^\varphi\{f \circledS g\} = \sqrt{j2\pi}\ \overline{F_S^\varphi}(-u_\varphi)\ G_S^\varphi(u_\varphi) \tag{56}$$

**Proof:** From the definition of SmFrFT (7), one obtains

$$\mathbb{T}_S^\varphi\{f \circledS g\} = \frac{1}{\sqrt{j2\pi}} \int_{-\infty}^{\infty} \{f(t) \circledS g(t)\} e^{-j\,t\,u_\varphi + \frac{j}{2}t^2\cot\varphi} dt$$

$$\mathbb{T}_S^\varphi\{f \circledS g\} = \frac{1}{\sqrt{j2\pi}} \int_{-\infty}^{\infty} \left\{\int_{-\infty}^{\infty} \bar{f}(\tau)\ g(t+\tau)\ W_{cr}(\tau,t)\ d\tau\right\} e^{-j\,t\,u_\varphi + \frac{j}{2}t^2\cot\varphi} dt \tag{57}$$

For solving (57), letting $t + \tau = \lambda$

$$\mathbb{T}_S^\varphi\{f \circledS g\} = \frac{1}{\sqrt{j2\pi}} \int_{-\infty}^{\infty} \int_{-\infty}^{\infty} \bar{f}(\tau)\ g(\lambda)\ e^{j\tau\lambda\cot\varphi} e^{-j\,(\lambda-\tau)\,u_\varphi + \frac{j}{2}(\lambda-\tau)^2\cot\varphi} d\tau\ d\zeta \tag{58}$$

Rearranging and noting that the complex-valued function $f(t) = f_1(t) + jf_2(t)$, one obtains

$$\mathbb{T}_S^\varphi\{f \circledS g\} = \frac{1}{\sqrt{j2\pi}} \int_{-\infty}^{\infty} \bar{f}(\tau)\ e^{ju_\varphi\tau + \frac{j}{2}\tau^2\cot\varphi} d\tau \times \sqrt{j2\pi}\ G_S^\varphi(u_\varphi) \tag{59}$$

where, $G_S^\varphi(u_\varphi) = \int_{-\infty}^{\infty} g(\lambda)\ e^{-ju_\varphi\lambda + \frac{j}{2}\lambda^2\cot\varphi} \tag{60}$

Solving (60) further, one obtains

$$\mathbb{T}_S^\varphi\{f \circledS g\} = \frac{1}{\sqrt{j2\pi}} \int_{-\infty}^{\infty} [f_1(\tau) - jf_2(\tau)]\ e^{ju_\varphi\tau + \frac{j}{2}\tau^2\cot\varphi} d\tau \times \sqrt{j2\pi}\ G_S^\varphi(u_\varphi)$$

$$\mathbb{T}_S^\varphi\{f \circledS g\} = \frac{1}{\sqrt{j2\pi}} \int_{-\infty}^{\infty} f_1(\tau)\ e^{-j(-u_\varphi)\tau + \frac{j}{2}\tau^2\cot\varphi} d\tau \times \sqrt{j2\pi}\ G_S^\varphi(u_\varphi) -$$

$$j\frac{1}{\sqrt{j2\pi}} \int_{-\infty}^{\infty} f_2(\tau)\ e^{-j(-u_\varphi)\tau + \frac{j}{2}\tau^2\cot\varphi} d\tau \times \sqrt{j2\pi}\ G_S^\varphi(u_\varphi) \tag{61}$$

Solving (61), one gets

$$\mathbb{T}_S^\varphi\{f \circledS g\} = \sqrt{j2\pi}\ F_{1S}^\varphi(-u_\varphi)\ G_S^\varphi(u_\varphi) - j\sqrt{j2\pi}\ F_{2S}^\varphi(-u_\varphi)\ G_S^\varphi(u_\varphi)$$

or, $\mathbb{T}_S^\varphi\{f \circledS g\} = \sqrt{j2\pi}\ \left[F_{1S}^\varphi(-u_\varphi) - jF_{2S}^\varphi(-u_\varphi)\right] G_S^\varphi(u_\varphi) = \sqrt{j2\pi}\ \overline{F_S^\varphi}(-u_\varphi)\ G_S^\varphi(u_\varphi) \tag{62}$

which proves the theorem.

***Special case:***

For the Euclidean FT, the rotation angle $\varphi = \pi/2$, then the expression (62) reduces to

$$\mathrm{T}_S^{\pi/2}\{f \circledcirc g\} = \sqrt{j2\pi}\ \overline{F_S^{\pi/2}}(-u_{\pi/2})\ G_S^{\pi/2}(u_{\pi/2})$$

$$\mathcal{F}\{f \circledcirc g\} = \sqrt{j2\pi}\ \overline{F_S}(-\omega)\ G_S(\omega) \tag{63}$$

This means that the proposed correlation theorem behaves similar to the Euclidean FT, except for the amplitude factor.

Some properties associated with the correlation theorem in SmFrFT domain are illustrated below.

***Property 1 (Shift Convolution).*** *Let $f$, $g \in L^1(\mathbb{R})$. The SmFrFT of $\mathbb{S}_d f \circledcirc g$ and $f \circledcirc \mathbb{S}_d g$ is given by*

$$\mathrm{T}_S^{\varphi}\{\mathbb{S}_d f \circledcirc g\}(u_{\varphi}) = \sqrt{j2\pi}\ e^{ju_{\varphi}d + \frac{j}{2}d^2 \cot\varphi}\ \overline{F_S^{\varphi}}(-u_{\varphi} - d\cot\varphi)\ G_S^{\varphi}(u_{\varphi}) \tag{64}$$

$$\mathrm{T}_S^{\varphi}\{f \circledcirc \mathbb{S}_d g\}(u_{\varphi}) = \sqrt{j2\pi}\ e^{-ju_{\varphi}d + \frac{j}{2}d^2 \cot\varphi}\ \overline{F_S^{\varphi}}(-u_{\varphi})\ G_S^{\varphi}(u_{\varphi} - d\cot\varphi) \tag{65}$$

*where, the symbol $\mathbb{S}_d$ represents the shift operator of a function by delay $d$ i.e., $\mathbb{S}_d x(t) = x(t - d), d \in \mathbb{R}$.*

***Proof:*** The shift correlation operator $\mathbb{S}_d f \circledcirc g$ is given by

$$(\mathbb{S}_d f \circledcirc g)(t) = \int_{-\infty}^{\infty} \bar{f}(\tau - d)\ g(t + \tau)\ W_{cr}(\tau, t)\ d\tau \tag{66}$$

where, $W_{cr}(\tau, t) = e^{j\tau(\tau + t)\cot\varphi}$. It implies

$$(\mathbb{S}_d f \circledcirc g)(t) = \int_{-\infty}^{\infty} \bar{f}(\tau - d)\ g(t + \tau)\ e^{j\tau(\tau + t)\cot\varphi}\ d\tau \tag{67}$$

Now, from the definition of SmFrFT (7), one obtains

$$\mathrm{T}_S^{\varphi}\{\mathbb{S}_d f \circledcirc g\}(u_{\varphi}) = \frac{1}{\sqrt{j2\pi}} \int_{-\infty}^{\infty} \{\mathbb{S}_d f(t) \circledcirc g(t)\}\ e^{-j\,t\,u_{\varphi} + \frac{j}{2}t^2 \cot\varphi}\ dt \tag{68}$$

Simplifying (68) further, one obtains

$$\mathbb{T}_S^{\varphi}\{\mathbb{S}_d f \circledast g\}(u_\varphi) = \frac{1}{\sqrt{j2\pi}} \int_{-\infty}^{\infty}\left\{\int_{-\infty}^{\infty} \bar{f}(\tau-d)\,g(t+\tau)\,e^{j\tau(\tau+t)\cot\varphi}\,d\tau\right\} e^{-j\,t\,u_\varphi + \frac{j}{2}t^2\cot\varphi}\,dt \tag{69}$$

To solve (69), lets assume $(t+\tau) = p$

$$\mathbb{T}_S^{\varphi}\{\mathbb{S}_d f \circledast g\}(u_\varphi) = \frac{1}{\sqrt{j2\pi}} \int_{-\infty}^{\infty} \bar{f}(\tau-d)\,e^{ju_\varphi\tau + \frac{j}{2}\tau^2\cot\varphi}\,d\tau \times \int_{-\infty}^{\infty} g(p)\,e^{-ju_\varphi p + \frac{j}{2}p^2\cot\varphi}\,dp \tag{70}$$

$$\mathbb{T}_S^{\varphi}\{\mathbb{S}_d f \circledast g\}(u_\varphi) = \int_{-\infty}^{\infty} \bar{f}(\tau-d)\,e^{ju_\varphi\tau + \frac{j}{2}\tau^2\cot\varphi}\,d\tau \times G_S^{\varphi}(u_\varphi) \tag{71}$$

Further, by letting $(\tau - d) = z$, and multiplying numerator and denominator of (71) by $\sqrt{j2\pi}$, one solves (71) as

$$\mathbb{T}_S^{\varphi}\{\mathbb{S}_d f \circledast g\}(u_\varphi) = \frac{1}{\sqrt{j2\pi}}\,e^{ju_\varphi d + \frac{j}{2}d^2\cot\varphi} \int_{-\infty}^{\infty} \bar{f}(z)\,e^{j(u_\varphi + d\cot\varphi)z + \frac{j}{2}z^2\cot\varphi}dz \times G_S^{\varphi}(u_\varphi) \times \sqrt{j2\pi} \tag{72}$$

Now, by letting the complex-valued function to be $f(z) = f_1(z) + jf_2(z)$, it implies $\bar{f}(z) = f_1(z) - jf_2(z)$ and thereby further reducing (72), one obtains

$$\mathbb{T}_S^{\varphi}\{\mathbb{S}_d f \circledast g\}(u_\varphi) = \sqrt{j2\pi}\,e^{ju_\varphi d + \frac{j}{2}d^2\cot\varphi} G_S^{\varphi}(u_\varphi)\left[\frac{1}{\sqrt{j2\pi}}\int_{-\infty}^{\infty} f_1(z)\,e^{j(u_\varphi + d\cot\varphi)z + \frac{j}{2}z^2\cot\varphi}dz - \right.$$
$$\left. j\frac{1}{\sqrt{j2\pi}}\int_{-\infty}^{\infty} f_1(z)\,e^{j(u_\varphi + d\cot\varphi)z + \frac{j}{2}z^2\cot\varphi}dz\right] \tag{73}$$

$$\mathbb{T}_S^{\varphi}\{\mathbb{S}_d f \circledast g\}(u_\varphi) = \sqrt{j2\pi}\,e^{ju_\varphi d + \frac{j}{2}d^2\cot\varphi} G_S^{\varphi}(u_\varphi)[F_{1S}^{\varphi}(-u_\varphi - d\cot\varphi) - jF_{2S}^{\varphi}(-u_\varphi - d\cot\varphi)]$$

$$\mathbb{T}_S^{\varphi}\{\mathbb{S}_d f \circledast g\}(u_\varphi) = \sqrt{j2\pi}\,e^{ju_\varphi d + \frac{j}{2}d^2\cot\varphi} G_S^{\varphi}(u_\varphi) \times \bar{F}_S^{\varphi}(-u_\varphi - d\cot\varphi) \tag{74}$$

Thus,

$$\mathbb{T}_S^{\varphi}\{\mathbb{S}_d f \circledast g\}(u_\varphi) = \sqrt{j2\pi}\,e^{ju_\varphi d + \frac{j}{2}d^2\cot\varphi} \bar{F}_S^{\varphi}(-u_\varphi - d\cot\varphi)\,G_S^{\varphi}(u_\varphi), \tag{75}$$

which proves the shift correlation property of the correlation theorem in SmFrFT domain.

Similarly, for solving $\mathbb{T}_S^\varphi \{ f \circledcirc \mathbb{S}_d g \}(u_\varphi)$ and utilizing the shift correlation operator of function $f \circledcirc \mathbb{S}_d g$ as $\int_{-\infty}^{\infty} \bar{f}(\tau) \, g(t + \tau - d) \, W_{cr}(\tau, t) \, d\tau$, where, $W_{cr}(\tau, t) = e^{j\tau(\tau+t)\cot\varphi}$ and based on the previous steps, one obtains

$$\mathbb{T}_S^\varphi \{ f \circledcirc \mathbb{S}_d g \}(u_\varphi) = \sqrt{j2\pi} \, e^{-ju_\varphi d + \frac{j}{2}d^2 \cot\varphi} \, \bar{F}_S^\varphi(-u_\varphi) \, G_S^\varphi(u_\varphi - d \cot\varphi), \tag{76}$$

which proves the shift correlation property of the correlation theorem in SmFrFT domain.

***Special case:***

For the FT, the rotation angle $\varphi = \pi/2$, then the expression () and () reduces to

$$\mathbb{T}_S^{\pi/2} \{ \mathbb{S}_d f \circledcirc g \}(u_{\pi/2}) = \sqrt{j2\pi} \, e^{-ju_{\pi/2} d} \, \bar{F}_S^{\pi/2}(-u_{\pi/2}) \, G_S^{\pi/2}(u_{\pi/2})$$

i.e, $\mathcal{F}\{ \mathbb{S}_d f \circledcirc g \}(\omega) = \sqrt{j2\pi} \, e^{-j\omega d} \, \bar{F}_S(-\omega) \, G_S(\omega)$ \hfill (77)

$$\mathbb{T}_S^{\pi/2} \{ f \circledcirc \mathbb{S}_d g \}(u_{\pi/2}) = \sqrt{j2\pi} \, e^{-ju_{\pi/2} d} \, \bar{F}_S^{\pi/2}(-u_{\pi/2}) \, G_S^{\pi/2}(u_{\pi/2})$$

i.e, $\mathcal{F}\{ f \circledcirc \mathbb{S}_d g \}(\omega) = \sqrt{j2\pi} \, e^{-j\omega d} \, \bar{F}_S(-\omega) \, G_S(\omega)$ \hfill (78)

This means that the proposed shift correlation property behaves similar to the Euclidean FT, as is evident from (77) and (78), except for the amplitude factor.

***Property 2*** (**Modulation Convolution**). *Let $f$, $g \in L^1(\mathbb{R})$. The SmFrFT of $\mathbb{M}_q f \circledcirc g$ and $f \circledcirc \mathbb{M}_q g$ is given by*

$$\mathbb{T}_S^\varphi \{ \mathbb{M}_q f \circledcirc g \}(u_\varphi) = \sqrt{j2\pi} \, \bar{F}_S^\varphi(-u_\varphi - q) \, G_S^\varphi(u_\varphi) \tag{79}$$

$$\mathbb{T}_S^\varphi \{ f \circledcirc \mathbb{M}_q g \}(u_\varphi) = \sqrt{j2\pi} \, \bar{F}_S^\varphi(-u_\varphi) \, G_S^\varphi(u_\varphi - q) \tag{80}$$

where, the symbol $\mathbb{M}_q$ represents the modulation operator, i.e., the modulation by $q$ of a function $x(t)$, $\mathbb{M}_q x(t) = e^{jqt} x(t), q \in \mathbb{R}$.

***Proof:*** The modulation convolution operator $\mathbb{M}_q f \circledcirc g$ is given by

$$\left(\mathbb{M}_q f \circledcirc g\right)(t) = \int_{-\infty}^{\infty} e^{jq\tau}\,\bar{f}(\tau)\,g(t+\tau)\,W_{cr}(\tau,t)\,d\tau \tag{81}$$

where, $W_{cr}(\tau,t) = e^{j\tau(\tau+t)\cot\varphi}$. It implies

$$\left(\mathbb{M}_q f \circledcirc g\right)(t) = \int_{-\infty}^{\infty} e^{jq\tau}\,\bar{f}(\tau)\,g(t+\tau)\,e^{j\tau(\tau+t)\cot\varphi}\,d\tau \tag{82}$$

Now, from the definition of SmFrFT (7), one obtains

$$\mathbb{T}_S^{\varphi}\{\mathbb{M}_q f \circledcirc g\}(u_\varphi) = \frac{1}{\sqrt{j2\pi}} \int_{-\infty}^{\infty} \{\mathbb{M}_q f(t) \circledcirc g(t)\}\, e^{-j\,t\,u_\varphi + \frac{j}{2}t^2\cot\varphi}\,dt \tag{83}$$

Simplifying (83) further, one obtains

$$\mathbb{T}_S^{\varphi}\{\mathbb{M}_q f \circledcirc g\}(u_\varphi) = \frac{1}{\sqrt{j2\pi}} \int_{-\infty}^{\infty}\int_{-\infty}^{\infty} \bar{f}(\tau)\,g(t+\tau)\ e^{jq\tau + j\tau(\tau+t)\cot\varphi - ju_\varphi t + \frac{j}{2}t^2\cot\varphi}\,d\tau\,dt \tag{84}$$

By letting $(t+\tau) = v$, (84) reduces to

$$\mathbb{T}_S^{\varphi}\{\mathbb{M}_q f \circledcirc g\}(u_\varphi) = \frac{1}{\sqrt{j2\pi}} \int_{-\infty}^{\infty} \bar{f}(\tau)\ e^{j(u_\varphi+q)\tau + \frac{j}{2}\tau^2\cot\varphi}\,d\tau \times \frac{1}{\sqrt{j2\pi}} \int_{-\infty}^{\infty} h(v)\ e^{-ju_\varphi v + \frac{j}{2}v^2\cot\varphi}\,dv \ \times \sqrt{j2\pi}$$

Simplifying further, one obtains

$$\mathbb{T}_S^{\varphi}\{\mathbb{M}_q f \circledcirc g\}(u_\varphi) = \sqrt{j2\pi}\ \bar{F}_S^{\varphi}(-u_\varphi - q)\ G_S^{\varphi}(u_\varphi), \tag{85}$$

which proves the modulation correlation property of the correlation theorem in SmFrFT domain.

Similarly, for solving $\mathbb{T}_S^{\varphi}\{f \circledcirc \mathbb{M}_q g\}(u_\varphi)$ and utilizing the modulation correlation operator of function $f \circledcirc \mathbb{M}_q g$ as $\int_{-\infty}^{\infty} \bar{f}(\tau)\,e^{jq(t+\tau)}\,g(t+\tau)\,W_{cr}(\tau,t)\,d\tau$, where, $W_{cr}(\tau,t) = e^{j\tau(\tau+t)\cot\varphi}$ and based on the previous steps, one obtains

$$\mathbb{T}_S^{\varphi}\{f \circledcirc \mathbb{M}_q g\}(u_\varphi) = \sqrt{j2\pi}\ \bar{F}_S^{\varphi}(-u_\varphi)\ G_S^{\varphi}(u_\varphi - q), \tag{86}$$

which proves the modulation correlation property of the correlation theorem in SmFrFT domain.

***Special case*:**

In case of FT, (85) and (86) reduces to (for $\varphi = \pi/2$)

$$\mathrm{T}_S^{\pi/2}\{\mathbb{M}_q f \circledcirc g\}(u_{\pi/2}) = \sqrt{j2\pi}\ \bar{F}_S^{\pi/2}(-u_{\pi/2}-q)\ G_S^{\pi/2}(u_{\pi/2}) = \sqrt{j2\pi}\ \bar{F}_S(-\omega-q)\ G_S(\omega),$$

i.e., $\mathcal{F}\{\mathbb{M}_q f \circledcirc g\}(\omega) = \sqrt{j2\pi}\ \bar{F}_S(-\omega-q)\ G_S(\omega)$ (87)

$$\mathrm{T}_S^{\pi/2}\{f \circledcirc \mathbb{M}_q g\}(u_{\pi/2}) = \sqrt{j2\pi}\ \bar{F}_S^{\pi/2}(-u_{\pi/2})\ G_S^{\pi/2}(u_{\pi/2}-q)$$

$$\mathcal{F}\{f \circledcirc \mathbb{M}_q g\}(\omega) = \sqrt{j2\pi}\ \bar{F}_S(-\omega)\ G_S(\omega-q)$$ (88)

This means that the proposed modulation correlation property behaves similar to the Euclidean FT, as is evident from (87) and (88), except for the amplitude factor.

***Property 3*** (**Time-Frequency shift Convolution**). *Let* $f$, $g \in L^1(\mathbb{R})$. *The SmFrFT of* $\mathbb{M}_q \mathbb{S}_d f \circledcirc g$ *and* $f \circledcirc \mathbb{M}_q \mathbb{S}_d g$ *is given by*

$$\mathrm{T}_S^{\varphi}\{\mathbb{M}_q \mathbb{S}_d f \circledcirc g\}(u_{\varphi}) = \sqrt{j2\pi}\ e^{-j(u_{\varphi}-q)d+\frac{j}{2}d^2\cot\varphi}\ \bar{F}_S^{\varphi}(u_{\varphi}-q-d\cot\varphi)\ G_S^{\varphi}(u_{\varphi})$$ (89)

$$\mathrm{T}_S^{\varphi}\{f \circledcirc \mathbb{M}_q \mathbb{S}_d g\}(u_{\varphi}) = \sqrt{j2\pi}\ e^{-j(u_{\varphi}-q)d+\frac{j}{2}d^2\cot\varphi}\ \bar{F}_S^{\varphi}(-u_{\varphi})\ G_S^{\varphi}(u_{\varphi}-q-d\cot\varphi)$$ (90)

*where, the symbol* $\mathbb{S}_d$ *and* $\mathbb{M}_q$ *represents the shift operator of a function by delay* $d$ *and the modulation operator of a function by* $q$, *i.e., for the function* $x(t)$, $\mathbb{S}_d x(t) = x(t-d)$, $d \in \mathbb{R}$ *and* $\mathbb{M}_q x(t) = e^{jqt}x(t)$, $q \in \mathbb{R}$.

***Proof:*** The time-frequency shift correlation operator is given by

$$(\mathbb{M}_q \mathbb{S}_d f \circledcirc g)(t) = \int_{-\infty}^{\infty} e^{jq\tau} \bar{f}(\tau-d)\ g(t+\tau)\ W_{cr}(\tau,t)\ d\tau$$ (91)

where, $W_{cr}(\tau,t) = e^{j\tau(\tau+t)\cot\varphi}$. It implies

$$(\mathbb{M}_q \mathbb{S}_d f \circledcirc g)(t) = \int_{-\infty}^{\infty} e^{jq\tau} \bar{f}(\tau-d)\ g(t+\tau)\ e^{j\tau(\tau+t)\cot\varphi}\ d\tau$$ (92)

The SmFrFT of (92) is obtained as

$$\mathrm{T}_S^{\varphi}\{\mathbb{M}_q \mathbb{S}_d f \circledcirc g\}(u_{\varphi}) = \frac{1}{\sqrt{j2\pi}} \int_{-\infty}^{\infty}\{\mathbb{M}_q \mathbb{S}_d f(t) \circledcirc g(t)\} e^{-j\ t\ u_{\varphi}+\frac{j}{2}t^2\cot\varphi} dt$$ (93)

Simplifying (93) further, one obtains

$$\mathbb{T}_S^{\varphi}\{\mathbb{M}_q\mathbb{S}_d\,f\circledcirc g\,\}(u_{\varphi}) = \frac{1}{\sqrt{j2\pi}}\int_{-\infty}^{\infty}\int_{-\infty}^{\infty}\bar{f}(\tau-d)\;g(t+\tau)\;e^{jq\tau+j\tau(\tau+t)\cot\varphi-j\,t\,u_{\varphi}+\frac{j}{2}t^2\cot\varphi}d\tau\,dt \qquad (94)$$

By letting $(t+\tau) = \varsigma$, (94) is simplified as

$$\mathbb{T}_S^{\varphi}\{\mathbb{M}_q\mathbb{S}_d\,f\circledcirc g\,\}(u_{\varphi}) = \frac{1}{\sqrt{j2\pi}}\int_{-\infty}^{\infty}\bar{f}(\tau-d)\;e^{jq\tau+ju_{\varphi}\tau+\frac{j}{2}\tau^2\cot\varphi}d\tau\times\frac{1}{\sqrt{j2\pi}}\int_{-\infty}^{\infty}g(\varsigma)\;e^{-ju_{\varphi}\varsigma+\frac{j}{2}\varsigma^2\cot\varphi}d\varsigma\times$$

$$\sqrt{j2\pi} \qquad (95)$$

Let $(\tau-d) = \xi$, (95) reduces to

$$\mathbb{T}_S^{\varphi}\{\mathbb{M}_q\mathbb{S}_d\,f\circledcirc g\,\}(u_{\varphi}) = \frac{1}{\sqrt{j2\pi}}\int_{-\infty}^{\infty}\bar{f}(\xi)\;e^{-j(u_{\varphi}-q-d\cot\varphi)\xi+\frac{j}{2}\xi^2\cot\varphi}d\xi\;\times\sqrt{j2\pi}\;\times\;e^{-ju_{\varphi}d+jqd+\frac{j}{2}d^2\cot\varphi}\times$$

$$G_S^{\varphi}(u_{\varphi}) \qquad (96)$$

Solving (96) further, by letting the complex-valued function $f(\xi) = f_1(\xi) + jf_2(\xi)$, one obtains

Thus, $\mathbb{T}_S^{\varphi}\{\mathbb{M}_q\mathbb{S}_d\,f\circledcirc g\}(u_{\varphi}) = \sqrt{j2\pi}\;e^{-j(u_{\varphi}-q)d+\frac{j}{2}d^2\cot\varphi}\;\bar{F}_S^{\varphi}(u_{\varphi}-q-d\cot\varphi)\;G_S^{\varphi}(u_{\varphi}),$ \qquad (97)

which proves the time-frequency shift correlation property of the correlation theorem in SmFrFT domain.

Similarly, for solving $\mathbb{T}_S^{\varphi}\{f\circledcirc\mathbb{M}_q\mathbb{S}_d g\}(u_{\varphi})$ and utilizing the shift and modulation correlation operator of function $f\circledcirc\mathbb{M}_q\mathbb{S}_d g$ as $\int_{-\infty}^{\infty}\bar{f}(\tau)\;e^{jq(t+\tau)}\;g(t+\tau-d)\;W_{cr}(\tau,t)\,d\tau$, where, $W_{cr}(\tau,t) = e^{j\tau(\tau+t)\cot\varphi}$ and based on the previous steps, one obtains

$$\mathbb{T}_S^{\varphi}\{f\circledcirc\mathbb{M}_q\mathbb{S}_d g\}(u_{\varphi}) = \sqrt{j2\pi}\;e^{-j(u_{\varphi}-q)d+\frac{j}{2}d^2\cot\varphi}\;\bar{F}_S^{\varphi}(-u_{\varphi})\;G_S^{\varphi}(u_{\varphi}-q-d\cot\varphi),$$ \qquad (98)

which proves the time-frequency shift correlation property of the correlation theorem in SmFrFT domain.

***Special case*:**

In case of FT, (97) and (98) reduces to (for $\varphi = \pi/2$)

$$\mathbb{T}_S^{\pi/2}\{\mathbb{M}_q\mathbb{S}_d\,f\circledcirc g\,\}(u_{\pi/2}) = \sqrt{j2\pi}\;e^{-j(u_{\pi/2}-q)d}\;\bar{F}_S^{\pi/2}(u_{\pi/2}-q)\;G_S^{\pi/2}(u_{\pi/2}),$$

i.e., $\mathcal{F}\{\mathbb{M}_q \mathbb{S}_d \, f \circledcirc g\,\}(\omega) = \sqrt{j2\pi}\, e^{-j(\omega-q)d}\, \bar{F}_S(\omega - q)\, G_S(\omega)$  (99)

$\mathbb{T}_S^{\pi/2}\{f \circledcirc \mathbb{M}_q \mathbb{S}_d \, g\,\}(u_{\pi/2}) = \sqrt{j2\pi}\, e^{-j(u_{\pi/2}-q)d}\, \bar{F}_S^{\pi/2}(-u_{\pi/2})\, G_S^{\pi/2}(u_{\pi/2} - q)$

i.e., $\mathcal{F}\{f \circledcirc \mathbb{M}_q \mathbb{S}_d \, g\,\}(\omega) = \sqrt{j2\pi}\, e^{-j(\omega-q)d}\, \bar{F}_S(-\omega)\, G_S(\omega - q)$  (100)

This means that the proposed time-frequency shift correlation property behaves similar to the Euclidean FT, as is evident from (99) and (100), except for the amplitude factor.

## 6. Conclusions and Future Scope of Work

In this paper, an elegant analytical expressions of convolution, product, and correlation of two functions is introduced in the simplified fractional Fourier transform domain. The newly established convolution, product, and correlation theorems along with their associated properties generalizes very nicely the classical result of Euclidean Fourier transform. The proposed approach offers the following advantages. First it is the first attempt to establish different signal processing theorems, which very nicely generalizes to the classical Fourier transform, which was not earlier attainable with the conventional definitions of fractional Fourier transform. Second, it has the added advantage of less computational complexity [14] as compared to the conventional fractional Fourier transform definitions, which will be beneficial for the reconfigurable implementation for different signal processing applications.

As a future work, the sampling of the bandlimited signals in the SmFrFT domain will be investigated based on the derived convolution theorem, with the establishment of the different formulae of uniform sampling and low pass reconstruction. Further, the approach of simplified fractional Fourier transform could be elaborated in the linear canonical transform and other angular transforms, which would prove to be an important mathematical tool for radar and sonar signal processing applications, along with the reconfigurable implementation for viable signal processing applications.

**Acknowledgements** The work was supported by Science and Engineering Research Board (SERB) (No. SB/S3/EECE/0149/2016), Department of Science and Technology (DST), Government of India, India.

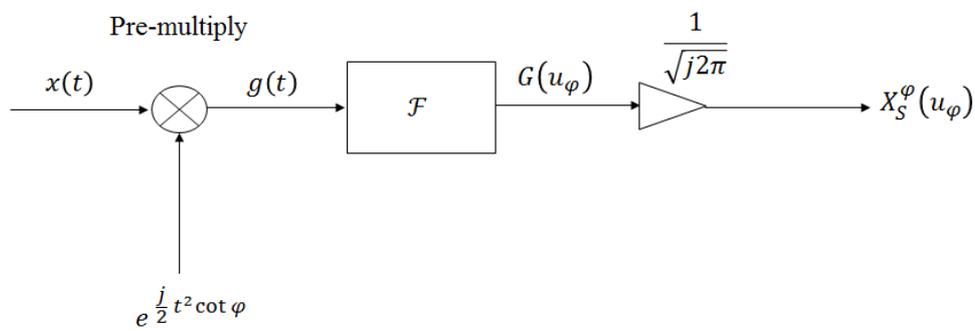

**Fig. 1.** Simplified fractional Fourier transform (SmFrFT) block diagram.